\documentclass[twocolumn]{aastex631}

\usepackage{color,soul}

\newcommand{\clusterfullname}{COOL\,J1241$+$2219}
\newcommand{\clustername}{COOL\,J1241}

\newcommand{\LENSTOOL}{{\tt{Lenstool}}}

\newcommand{\zarc}{5.043}
\newcommand{\zcluster}{1.001}

\newcommand{\modelNpar}{14}
\newcommand{\modelNconstraints}{24}
\newcommand{\modelrms}{$0\farcs16$} 
\newcommand{\massinfhkpc}{$M(< 500 $kpc$) = 3.6^{+0.8}_{-0.4}\times 10^{14} M_{\sun}$}
\newcommand{\massinRarc}{$M(<5\farcs77) =1.079^{+0.023}_{-0.007}\times 10^{13} M_{\sun}$}
\newcommand{\magnificationci}{$6.6 ^{+2.6}_{-2.2}$} 
\newcommand{\magnificationarc}{$76 ^{+40}_{-20}$} 

\newcommand{\magnificationarcw}{$92^{+37}_{-31}$}

\newcommand{\SFRn}{$11.4^{+7.4}_{-5.9}$}
\newcommand{\LMStarn}{$9.8\pm0.3$}
\newcommand{\SFRfw}{$10.3^{+7.0}_{-4.4}$}
\newcommand{\LMStarfw}{$9.7\pm0.3$}

\begin{document}

\defcitealias{Khullar_2021}{K21}

\title{COOL-LAMPS VI: Lens model and New Constraints on the Properties of COOL\,J1241$+$2219, a Bright $z=5$ Lyman Break Galaxy and its $z=1$ Cluster Lens}


\author[0000-0001-5786-1272]{Maxwell Klein}
\affiliation{Department of Astronomy, University of Michigan, 1085 S. University Ave, Ann Arbor, MI 48109, USA}

\author[0000-0002-7559-0864]{Keren Sharon}
\affiliation{Department of Astronomy, University of Michigan, 1085 S. University Ave, Ann Arbor, MI 48109, USA}
 
\author[0000-0003-4470-1696]{Kate Napier}
\affiliation{Department of Astronomy, University of Michigan, 1085 S. University Ave, Ann Arbor, MI 48109, USA}

\author[0000-0003-1370-5010]{Michael D. Gladders}
\affiliation{Department of Astronomy and Astrophysics, University of Chicago,
 5640 South Ellis Avenue, Chicago,
 IL  60637, USA}
\affiliation{Kavli Institute for Cosmological Physics,
University of Chicago, Chicago, IL 60637, USA}

\author[0000-0002-3475-7648]{Gourav Khullar}
\affiliation{Department of Physics and Astronomy, and PITT PACC, University of Pittsburgh, Pittsburgh, PA 15260, USA}

\author[0000-0003-1074-4807]{Matthew Bayliss}
\affiliation{Department of Physics, University of Cincinnati, Cincinnati, OH 45221, USA}

\author[0000-0003-2200-5606]{H{\aa}kon Dahle}
\affiliation{Institute of Theoretical Astrophysics, University of Oslo, P.O. Box 1029, Blindern, NO-0315 Oslo, Norway}

\author[0000-0002-2862-307X]{M. Riley Owens}
\affiliation{Department of Physics, University of Cincinnati, Cincinnati, OH 45221, USA}

\author[0000-0002-2718-9996]{Antony Stark}
\affiliation{Center for Astrophysics $|$ Harvard \& Smithsonian
Observatory Building E, 60 Garden St, Cambridge, MA 02138}

\author[0000-0002-5430-4355]{Sasha Brownsberger}
\affiliation{Department of Physics, Harvard University, 17 Oxford Street, Cambridge, MA 02138, USA}

\author[0000-0001-6505-0293]{Keunho J. Kim}
\affiliation{Department of Physics, University of Cincinnati, Cincinnati, OH 45221, USA}

\author[0000-0002-7072-7622]{Nicole Kuchta}
\affiliation{Department of Astronomy, University of Michigan, 1085 S. University Ave, Ann Arbor, MI 48109, USA}
\author[0000-0003-3266-2001]{Guillaume Mahler}
\affiliation{Centre for Extragalactic Astronomy, Durham University,
South Road, Durham DH1 3LE, UK 2}
\affiliation{Institute for Computational Cosmology, Durham University, South Road, Durham DH1 3LE, UK}
\author[0000-0003-3971-5727]{Grace Smith}
\affiliation{Department of Astronomy, University of Michigan, 1085 S. University Ave, Ann Arbor, MI 48109, USA}
\author[0000-0001-5424-3698]{Ryan Walker}
\affiliation{Department of Astronomy, University of Michigan, 1085 S. University Ave, Ann Arbor, MI 48109, USA}

\author[0000-0003-2294-4187]{Katya Gozman}
\affiliation{Department of Astronomy, University of Michigan, 1085 S. University Ave, Ann Arbor, MI 48109, USA}

\author[0000-0002-8397-8412]{Michael N. Martinez}
\affiliation{Department of Physics, University of Wisconsin-Madison, 1150 University Avenue, Madison, WI, 53706, USA}

\author[0000-0001-9225-972X]{Owen S. Matthews Acu\~{n}a}
\affiliation{Department of Astronomy, University of Wisconsin–Madison, 475 N Charter St, Madison, WI, 53706, USA}

\author[0000-0001-5931-5056]{Kaiya Merz}
\affiliation{Space Telescope Science Institute, 3700 San Martin Drive, Baltimore, MD 21218, USA}

\author[0000-0002-9142-6378]{Jorge A. Sanchez}
\affiliation{School Of Earth and Space Exploration, 781 Terrace Mall, Tempe, AZ 85287, USA}

\author[0000-0001-8008-7270]{Daniel J. Kavin Stein}
\affiliation{Department of Astronomy and Astrophysics, University of
Chicago, 5640 South Ellis Avenue, Chicago, IL 60637, USA}

\author[0000-0002-1106-4881]{Ezra O. Sukay}
\affiliation{Department of Physics and Astronomy, Johns Hopkins University, 3400 North Charles Street, Baltimore, MD 21218, USA.}

\author[0000-0001-6584-6144]{Kiyan Tavangar}
\affiliation{Department of Astronomy, Columbia University, 550 West 120th Street, New York, NY, 10027, USA}

\correspondingauthor{Keren Sharon}
\email{kerens@umich.edu}

\begin{abstract}
We present a strong lensing analysis of COOL\,J1241$+$2219, the brightest known gravitationally lensed galaxy at  $z \geq 5$, based on new multi-band Hubble Space Telescope (HST) imaging data. 
The lensed galaxy has a redshift of $z = 5.043$, placing it shortly after the end of the Epoch of Reionization, and an AB magnitude $z_{\rm AB}=20.47$ mag \citep{Khullar_2021}. 
As such, it serves as a touchstone for future research of that epoch. 
The high spatial resolution of HST reveals internal structure in the giant arc, from which we identify 15 constraints and construct a robust lens model. We use the lens model to extract cluster mass and lensing magnification.  We find that the mass enclosed within the Einstein radius of the $z = 1.001$ cluster lens is \massinRarc, significantly lower than other known strong lensing clusters at its redshift. 
The average magnification of the giant arc is $<\mu_{arc}>=$\magnificationarc, a factor of $2.4^{+1.4}_{-0.7}$ greater than previously estimated from ground-based data; the flux-weighted average magnification is $<\mu_{arc}>=$\magnificationarcw. We update the current measurements of the  stellar mass and star formation rate (SFR) of the source for the revised magnification, $\log(M_\star/M_{\odot}) = $ \LMStarfw\ and ${\rm SFR} = $\SFRfw $ M_{\odot} $yr$^{-1}$. 
The powerful lensing magnification acting upon COOL\,J1241$+$2219 resolves the source and enables future studies of the properties of its star formation on a clump-by-clump basis. The lensing analysis presented here will support upcoming multiwavelength characterization with HST and JWST data of the stellar mass assembly and physical properties of this high-redshift lensed galaxy.

\end{abstract}


\section{Introduction} \label{sec:intro}

Gravitational lensing is nature’s telescope: massive foreground structures, such as galaxy clusters, magnify and distort background galaxies, greatly increasing the angular extent of the source and the number of photons that reach the observer. Using observational evidence of gravitational lensing, we can reconstruct the gravitational potential of the lens and study the mass distribution of the foreground cluster and the physical properties of the distant galaxies that the cluster lenses. Strong gravitational lensing can greatly magnify galaxies that would otherwise be too dim to study, allowing for spatial resolution of 100s or even 10s of parsecs \citep[e.g.,][]{Johnson_2017Rec,Johnson_2017,Cornachione_2018,Claeyssens_2023,Welch_2023,Vanzella_2023}. With a robust lens model, the lensed galaxy can be de-lensed into a high-resolution projection of the background galaxy in the source plane. These fortuitous alignments of background galaxies and strong gravitational lenses give us the best possible glimpse into the distant universe, providing detailed information on the physical properties of galaxies.
While wide surveys like the Sloan Digital Sky Survey \cite[SDSS;][]{Ahumada_2020} have identified many bright $z < 3$ lensed galaxies, such discoveries are rare at higher redshifts \citep{Franx_1997, Soifer_1998,Frye_2002,Kubo_2010,Bayliss_2011,Stark_2013,Gladdersz488}. 

\clusterfullname\ (hereafter \clustername), is a bright, strongly lensed galaxy at $z=$ \zarc, lensed by a galaxy cluster at $z=$\zcluster. It was discovered by the ChicagO Optically selected Lenses---Located At the Margins of Public Surveys collaboration \citep[COOL-LAMPS;][hereafter, \citetalias{Khullar_2021}]{Khullar_2021}. The program seeks to find strong lensing systems in sky surveys for follow-up observation and analysis. The discovery, and an initial analysis based on ground-based data, are reported in \citetalias{Khullar_2021}.
Remarkably, \clustername\ has an AB magnitude $z = 20.47$ mag; it is $\sim$ 5 to 13 times brighter and $2-4\%$ farther away than the brightest known lensed galaxies at $z\sim 5$ \citep{Gladdersz488,Franx_1997,Soifer_1998,Frye_2002}. 
\clustername\ is one of the rarest and most fortunate examples of strong gravitational lensing. 

By the redshift of \clustername, $z \sim 5$, the Epoch of Reionization had ended (e.g., \citealt{Fan06}; see \citealt{Gnedin_2022, Robertson_2022} for recent reviews) and the cosmic star formation rate was rapidly increasing.  
\citetalias{Khullar_2021}\ found that \clustername\ has a stellar mass similar to that of the Milky Way and a star formation rate (SFR) that is one to two orders of magnitude higher. However, \clustername\ is unlike other known galaxies within this epoch due to its lower relative dust content and SFR \citepalias{Khullar_2021}. This points to a star formation history with potentially significant mass assembly in the recent past (for it to have formed its stellar mass so rapidly); although how many epochs of bursty star formation occurred, or how long these star forming episodes were, is unclear with extant data. The effect of dust-enshrouded star formation on total star formation rate in this epoch is still not understood \citep{Casey_2019,Mowla_2022,Claeyssens_2023}. Upcoming JWST data (JWST Cycle 1, ID 2566; PI: Khullar; \citealt{Khullar2021jwstprop}) will allow for a deconstruction of the complete star formation history and characterization of mass assembly in this system.

\citetalias{Khullar_2021}\ constructed a lens model of this system using ground-based data in the $J-$ and $H-$ bands from the FourStar Infrared Camera \citep[FourStar;][]{Persson_2008} on the Magellan/Baade telescope in Chile, as well as in the $g$, $r$, $i$, and $z$ bands from the Parallel Imager for Southern Cosmology Observations \citep[PISCO;][]{Stalder_2014}. As is the case with ground-based data, the resolution was low. Nevertheless, the data still enabled a rudimentary lens model and mass approximation for the galaxy cluster, and a preliminary source analysis. With HST, a higher spatial resolution provides better lensing constraints, leading to a more robust mass and magnification measurement \citep[e.g.,][]{Keren_2012Rec,Sukay2022,Zhuang2023}.

The goal of this paper is to present an improved, space-based lens model that allows us to extract more accurate and precise information about both the lens and the source, and create a stepping stone towards future study with forthcoming JWST data. 
This paper is structured as follows. In Section \ref{sec:Data}, we present the HST imaging and data reduction. In Section \ref{sec:Methods}, we present the methods used to generate a lens model, measure mass and magnifications, and obtain their uncertainties. In Section \ref{sec:results}, we present the results of the model, namely the mass and magnification maps, as well as the magnification measurements. Finally, in Section \ref{sec:Discussion}, we use our results to revise the findings of \citetalias{Khullar_2021}, and discuss the significance of the results in the context of high redshift galaxies and strong lensing galaxy clusters at $z \gtrsim 1 $.

We assume a flat universe with parameters $H_0 = 70$ km s$^{-1}$ Mpc$^{-1}$, $\Omega_{m}$ = 0.3, and $\Omega_{\Lambda} = 0.7$. Magnitudes are presented in the AB system. The spectroscopically measured cluster and source redshifts are $z_{cluster} = 1.001$, $z_{source}=\zarc$, respectively, which correspond to $8.0106$ and $6.2561$ kpc per arcsec in this cosmology. 

\begin{figure*}[!t]
    \centering
    \includegraphics[width=\textwidth, height=\textheight, keepaspectratio]{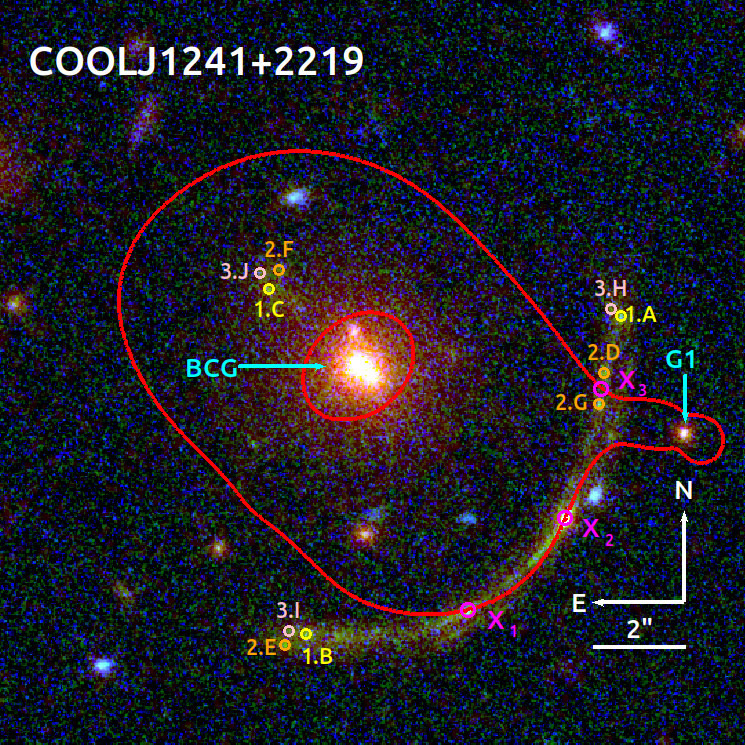}
    \caption{HST imaging of \clustername, rendered from F160W (red), F814W (green) and F606W (blue). The best-fit lens model is overplotted. Solid red lines denote the critical curve for a source at $z=$ \zarc. The cyan arrows point to the brightest cluster galaxy (BCG), and a galaxy-scale halo (G1) whose parameters were optimized by the model. Magenta $\times$ symbols mark the locations of critical curve constraints (X$_2$ was carried over from \citetalias{Khullar_2021}, and X$_2$ was constraint 1.2 in \citetalias{Khullar_2021}). 
    Multiple images that were used to constrain the lens model are labeled with circles and letters, and color-coded according to their source identification (see \autoref{tab:coords}).}
    \label{fig:generalmodel}
\end{figure*}

\section{Data} \label{sec:Data}

The data were collected as part of a joint Very Large Array (VLA) and HST program (VLA/2020-07-025, HST-GO 16484, PI: A. Stark; \citealt{stark2020hstprop}). The data include 3 orbits of HST imaging split between four broad-band filters: half an orbit each for two WFC3/IR bands, F110W (1012 s) and F160W (1212 s) observed on 2022 Jan 4, and one orbit in each of two ACS bands, F814W and F606W (2080 s each) observed on 2021 Dec 10.  The ACS data were obtained with a simple 4-point dither sequence using the standard \texttt{ACS-WFC-DITHER-BOX}. Because the observed lensing configuration suggested that a few tens of arcseconds around the lens center are of most interest, there was no need to fill the chip gap. The chip 2 pointing aperture was used in order to avoid a star on the field edges that could enter the field of view at some roll angles.

The orbit packing of the single WFC3-IR orbit was guided by a goal of obtaining a minimum of four dithered images per filter for adequate PSF sampling. Four images per filter were obtained with the \texttt{WFC3-IR-DITHER-BOX-MIN} pattern, using \texttt{NSAMP}=11 and \texttt{SPARS25} for F110W and \texttt{NSAMP}=7 and \texttt{SPARS50} for F160W, resulting in a slightly longer observing time in F160W due to buffer time constraints. 

We reduced the data using the DrizzlePac python package\footnote{\url{http://www.stsci.edu/scientific-community/software/drizzlepac.html}} as follows.
We obtained the individual pre-processed frames from the MAST archive shortly after data acquisition. We used \texttt{AstroDrizzle} to co-add the sub-exposures of each filter, using a gaussian kernel drop size of $0.8$ for the ACS images and $1.0$ for WFC3. For WFC3/IR data, we applied a secondary correction of the infrared (IR) blobs using a model-based flat-fielding technique \citep{Sharon_2020} prior to combining the individual IR exposures.
We used \texttt{TweakReg} to match the WCS solution of imaging data that were obtained in different epochs (i.e., the WFC3 and ACS data). We then propagated the solution back to the individual frames using \texttt{TweakBack}, and re-drizzled to obtain the final set of CTE corrected, cosmic-ray and bad-pixel rejected images, with a common reference frame and a pixel scale of $0\farcs03~\rm{pixel}^{-1}$.
A representative HST (F160W, F814W, F606W) color image of the strong lensing region of \clustername\ is shown in \autoref{fig:generalmodel}.

\section{Methods} \label{sec:Methods}
\subsection{Lensing Evidence} \label{sec:Lensing Constraints}
The lensing analysis of \citetalias{Khullar_2021}\ assumed that the lensing configuration of \clustername\ is a common three-image arc with a counter-image, and that a high magnification near the critical curve is responsible for the high brightness of a clump near the center of the giant arc. As we describe below, the ground-based imaging used for their analysis
had insufficient resolution to reveal the complexity of the system.

In the common three-image arc configuration, three near-complete images of the source merge to form a giant arc (e.g., SDSS\,J1110$+$6459, \citealt{Johnson_2017}; SPT-CL\,J2011$-$5228, \citealt{Collett_2017}) with the critical curve crossing the arc in two places, defining clear mirror symmetry on its opposite sides. If a counter-image forms (in configurations other than a ``naked cusp''), it is usually a less-magnified, complete image.
The high resolution of our shallow HST imaging reveals the clumpiness of the source, and makes it clear that the lensing configuration of \clustername\ deviates from the common three-image arc configuration.   
We identified several star-forming clumps along the arc and used their relative surface brightness and slight variations in color to match and identify them as multiple image instances of unique regions in the source galaxy. We label the clumps in \autoref{fig:generalmodel}, and their coordinates are listed in \autoref{tab:coords}. 
We find that similarly to the common three-image arc configuration described above, \clustername\ has an image of the source in each end of its giant arc, which match the complete, less magnified counter image near the brightest cluster galaxy (BCG). However, most of the middle segment of the arc is not easily mapped to the other images. It appears to contain two bright clumps that do not match other clumps in terms of surface brightness and morphology. The counter image lacks a clump with such high surface brightness. The only likely explanation of these bright clumps along the arc is that they are a result of extreme magnification, caused by very close proximity to the source plane caustic (a.k.a., ``caustic crossing''), as was concluded by \citetalias{Khullar_2021}\ for the brightest clump. Such scenarios have been reported in the literature for other systems \citep[e.g.,][]{Welch2022hst,Welch2022jwst,Diego2022,Meena2023}. In the case of \clustername, it is likely that the central portion of the arc forms along the critical curve like a segment of an Einstein ring \citep[e.g., the ``Cosmic Horseshoe'';][]{Bellagamba2017}.

\citetalias{Khullar_2021}\ used as constraints three brightness peaks along the arc, under the assumption that these are the cores of the lensed galaxy, matched to the center of the candidate counter image. In addition, they used two regions with similar surface brightness marking the edges of the star-forming clump that was assumed to be merging on the critical curve. 
 
With the new high-resolution HST imaging, we find that the brightest regions of the giant arc formed as a result of high magnification or multiplicity rather than being a dominant core of the source galaxy. 
For the brightest of the clumps, our analysis concurs with the assumption of \citetalias{Khullar_2021}\ that the brightest clump sits on top of the critical curve. Conversely, the peak brightness in the north segment of the arc, identified as 1.3 in \citetalias{Khullar_2021}, is actually resolved in the HST data into two images of one star-forming clump separated by $0\farcs7$, with mirror symmetry (labeled 2.D and 2.G in \autoref{fig:generalmodel} and \autoref{tab:coords}). To form these two images requires that a critical curve crosses the arc north of its assumed crossing in the \citetalias{Khullar_2021} model, likely due to a contribution from a cluster-member galaxy $\sim1\farcs7$ west of the arc that complicates the lensing potential in that region.

Image 1.2 of \citetalias{Khullar_2021}\ is a compact clump whose brightness and morphology are not matched by any other clump, and therefore could also be associated with caustic crossing. 
Finally, the regions north and south of the brightest clump are not symmetrical, providing evidence that the critical curve is not perpendicular to the arc at this location. We describe the new lensing constraints below.

Our model uses as constraints three unique star-forming clumps within the source galaxy of \clustername, and three critical curve constraints on the giant arc. The clumps were identified based on similarities in surface brightness and color. Only the most secure clump identifications were used as constraints. 
We did not include constraints from the middle section of the arc because these were hard to robustly identify as multiple images of the clumps in the north and south ends of the arc and the counter image. This region is primarily constrained by forcing the critical curve to pass through points on the arc as determined by the lensing symmetry.

The critical curve constraint $X_2$ marks the brightest part of the giant arc, a constraint that was carried over from \citetalias{Khullar_2021}. $X_3$ in the north of the arc forces the model to form the mirror image geometry of 2.D and 2.G. Finally, $X_1$ accounts for a bright clump that does not map properly to other images. The most likely interpretation is that a region of the background galaxy is located almost directly under the lensing caustic, resulting in a highly magnified image of that region. Adding this constraint to the model resulted in better agreement between predicted and observed lensing constraints, i.e., better image-plane rms.
In \autoref{tab:coords}, we list the coordinates, group, and image IDs of each of the identified image constraints, and the location of the critical curve constraints in arcseconds relative to the BCG.

While the HST data resulted in several precise constraints along the giant arc and its counter-image, which improved upon the constraints of \citetalias{Khullar_2021}, the data did not reveal any new lensed galaxies. We detected some blue sources in the field, but were unable to robustly confirm them as strongly lensed nor obtain a spectroscopic redshift for them. Additionally, the location of some of these candidates, whose projected distance from the BCG placed them farther than the radius of the main arc, implies that they should be at a higher redshift than \clustername. Such a  high redshift is ruled out by the blue color of these candidates. We, therefore, conclude that these sources are unlikely strongly lensed and exclude them from our analysis.

The lack of constraints from multiple source planes is a weakness of the model, as its leverage over the inner slope of the mass distribution is limited, resulting in systematic uncertainties that are not well captured by the error analysis \citep[e.g.,][]{johnson2016}. 

\subsection{Lens Modeling}

We computed the model using the \LENSTOOL\ \citep{Jullo_2007} software, following the procedure described in \cite{Sharon_2020}. \LENSTOOL\ assumes a parametric model of the projected mass distribution of the cluster lens, and uses Markov Chain Monte Carlo (MCMC) sampling of the parameter space to find the model that results in the smallest image-plane rms scatter between the observed images and those predicted by the lens model. The cluster lens was modeled as a linear combination of mass halos representing the dark matter distribution of the cluster and its large scale structure as well as individual cluster-member galaxies. 
Each halo was assumed to have a pseudo-isothermal ellipsoidal mass distribution \cite[dPIE, also referred to as PIEMD in the literature;][]{Jullo_2007, eliasdottir2007} with seven parameters: position ($x$,  $y$), ellipticity ($e$), position angle ($\theta$), core and cut radii ($r_{core}$, $r_{cut}$), and effective velocity dispersion ($\sigma$).
Cluster-member galaxies were selected photometrically, based on their color with respect to the red sequence \citep{Gladders_2000} in a color-magnitude diagram. We used F814W-F110W vs F110W, which provides a good sampling of the 4000 $\mathrm{\AA}$ break in the spectral energy distribution of quiescent galaxies at the cluster redshift. The cluster-member galaxies were also parameterized as dPIE halos, with positional parameters ($x$, $y$, $\theta$, $e$) fixed to their light distribution as measured with Source Extractor \citep{bertin1996} and their slope parameters and normalization scaled to their measured magnitudes in F110W using scaling relations as described in \citet{Limousin2005}.

The best-fit solution of the lens plane used three halos with free parameters: a cluster-scale halo representing the cluster, and two galaxy-scale halos that were decoupled from the cluster-member catalog. These cluster-member galaxies were the BCG, which is not expected to follow the same scaling relations as other cluster-member galaxies; and the galaxy labeled G1 in \autoref{fig:generalmodel}, because of its proximity to the giant arc. The positional parameters of these galaxies were fixed, but some or all of their slope parameters were allowed to vary and the best-fit solution was identified by the lens modeling process.

\begin{table}[h!]
    \centering
    \begin{tabular}{||c|c|c|c|c|} 
    \hline
         Source & Image & R.A. (J2000) & Decl. (J2000) & $\mu$\\
                &       & (deg)        & (deg)         &  \\
    \hline
    1 &A &190.3755783 &22.3297457 &$33^{+17}_{-8}$\\
    1 &B &190.3755783 &22.3278230 &$17^{+9}_{-5}$\\
    1 &C &190.3735237 &22.3299101 &$6^{+3}_{-2}$\\
    2 &D &190.3736331 &22.3294010 &$275^{+416}_{-139}$\\
    2 &E &190.3757184 &22.3277581 &$14^{+8}_{-4}$\\
    2 &F &190.3757558 &22.3300264 &$6^{+4}_{-2}$\\
    2 &G &190.3736638 &22.3292139 &$179^{+313}_{-48}$\\
    3 &H &190.3758224 &22.3297899 &$35^{+17}_{-9}$\\
    3 &I &190.3735880 &22.3300055 &$16^{+9}_{-4}$\\
    3 &J &190.3758824 &22.3278459 &$7^{+4}_{-2}$\\
    \hline
    Crit &$X_1$ &190.3745218 &$22.3279684$&\nodata\\
    Crit &$X_2$ &190.3738858 &$22.3285245$&\nodata\\
    Crit &$X_3$ &190.3736529 &$22.3293070$&\nodata\\
    \hline
    \end{tabular}
    \caption{Coordinate location of constraints and the corresponding magnifications, as derived from the best-fit model. Uncertainties are inferred from the MCMC sampling of the parameter space and represent the difference between the best-fit value and the 84th and 16th percentile. The critical curves are constrained to pass within 0.05" of the listed coordinate}
    \label{tab:coords}
\end{table}

\begin{deluxetable*}{clccccccc} 
\tablecolumns{9} 
\tablehead{\colhead{No. }   & 
            \colhead{Component }   & 
            \colhead{$\Delta$ R.A. ($\arcsec$)}     & 
            \colhead{$\Delta$ Decl. ($\arcsec$)}    & 
            \colhead{$e$}    & 
            \colhead{$\theta$ (deg)}       & 
            \colhead{$r_{\rm core} $ (kpc)} &  
            \colhead{$r_{\rm cut}$ (kpc)}  &  
            \colhead{$\sigma_0$ (km s$^{-1}$)}             } 
\startdata 
1 & Cluster & $-0.62_{-1.92}^{+4.21}$ & $-2.51_{-1.37}^{+4.10}$ & $0.18_{-0.10}^{+0.06}$ & $131_{-9}^{+3}$ & $133_{-51}^{+66}$ & [1500] & $985_{-222}^{+262}$ \\
2 & BCG &[0.0] & [0.0]  & $0.07_{-0.02}^{+0.23}$& $140_{-23}^{+15}$  & $4.5_{-4.4}^{+0.3}$ & $50_{-25}^{+47}$  & $298_{-86}^{+2}$   \\
3 & G1 &[$-7.078359$] & [$-1.463040$] &[0.033] & [$-81.44$] & [0.029] & [9.5]  & $95\pm{26}$\\
&L* galaxy  & \nodata & \nodata & \nodata & \nodata & \nodata  &    $10.7_{-6.3}^{+63.5}$ &  $137_{-80}^{+8}$ \\ 
\enddata 

\caption{Best-fit lens model parameters. Coordinates are tabulated in arcseconds from the center of the BCG, [R.A., decl.]=[$190.3752375$, $22.3294502$].  All the mass components were parameterized as dPIE (see \autoref{sec:Methods}), with ellipticity expressed as $e=(a^2-b^2)/(a^2+b^2)$. $\theta$ is measured North of West. Statistical uncertainties were inferred from the MCMC optimization and correspond to a 95$\%$ confidence interval. Parameters in square brackets were not optimized. The location and the ellipticity of cluster galaxies were kept fixed according to their light distribution, and the other parameters were determined through scaling relations (see text). The image plane rms of the best-fit model is \modelrms.}\label{tab:model}
\end{deluxetable*} 

\begin{figure}[!t]
    \centering
     \includegraphics[width=0.47\textwidth, keepaspectratio]{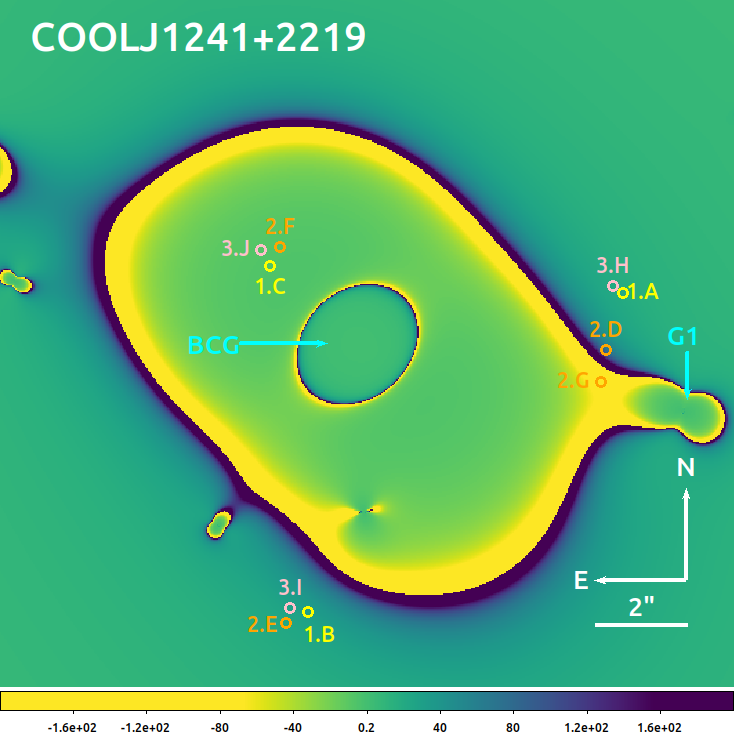}
    \caption{The magnification map of the galaxy cluster, centered on the BCG, for a source at $z=$\zarc. Colorscale indicates magnification, with positive and negative values having opposite lensing parity. The field of view and labels are the same as \autoref{fig:generalmodel}.}
    \label{fig:generalmag}
\end{figure}

\section{Results} \label{sec:results}
The lens plane is well described by a dominant dark-matter halo, representing the cluster-scale mass distribution, supplemented with galaxy-scale halos representing the BCG and cluster-member galaxies. Contribution to the lensing potential from galaxy G1 was required in order to reproduce the lensing configuration of the star forming clumps observed in the giant arc. The normalization parameter $\sigma$ of G1 was optimized individually, in order to add flexibility to the lens model and avoid biasing the scaling relation parameters towards a solution that is primarily driven by one galaxy. 
The model predicts a demagnified fifth image of the source behind the BCG, 5-6 magnitudes fainter than the observed arcs. Given its magnification, it is not expected to be visible in the data.
The model has \modelNpar\ free parameters and \modelNconstraints\ constraints and results with an rms scatter of \modelrms\ for predicted images in the image plane. 

In \autoref{tab:model} we tabulate the best-fit lens model parameters and their uncertainties, as determined from the MCMC sampling of the parameter space.
\autoref{fig:generalmodel} displays the lensing constraints and the critical curves for a source at $z=\zarc$ derived from the best-fit lens model.

\autoref{fig:generalmag} shows the best-fit magnification map for a source at $z=\zarc$.  Representative magnification values at the locations of the lensing constraints are presented in \autoref{tab:coords}. The errors were estimated by generating 100 models sampled from the MCMC chain and finding the 16th and 84th percentile of the magnification at each coordinate. The uncertainties represent statistical modeling errors, and do not take into account systematic uncertainties. As expected, regions in close proximity to the critical curve have higher magnification and uncertainty.

A calculation of the average magnification of the entire arc is complicated by the proximity of the arc to the critical curve, which marks regions in the image plane where a point source would theoretically have infinite magnification. To avoid numerical effects near the critical curve, we calculated the magnification by comparing the image plane area to the source plane area as follows. We traced the arc with an aperture of width $0\farcs8$; using the deflection matrices $\alpha$ of the best-fit model and the 100 models sampled from the MCMC, we ray-traced the aperture to the source plane using the lens equation, $\beta = \theta - \alpha(\theta)$, where $\beta$ is the source position, $\theta$ is the image position, and $\alpha$ is the lensing deflection at position $\theta$. We then used the ratio of image plane to source plane areas to calculate the average magnification and its uncertainty. The same approach was used by \citetalias{Khullar_2021}; we compare our magnification measurements in \autoref{sec:Discussion}. 
The average magnification of the entire giant arc is $<\mu_{arc}>=$ \magnificationarc, and the counter image is magnified by $<\mu_{c.i.}>=$ \magnificationci.

The mean projected distance between the giant arc and the BCG is $5\farcs77$. The mass enclosed within this radius, from our lensing analysis, is \massinRarc. The analytical expression for the mass enclosed within the Einstein radius of a spherically symmetrical lens is $M(<\theta_E)=\Sigma_{cr}\pi[D_L \theta_E]^2= 1.17 \times 10^{13} M_{\odot}$ for an Einstein radius of $\theta_E=5\farcs77$, corresponding to 46.2 kpc at the lens redshift. In this equation, $\Sigma_{cr}$ is the critical density for lensing, and $D_L$ is the angular diameter distance to the lens \citep[see, e.g.,][]{kneib2011}. Using numerical simulations, \cite{Remolina2020} showed that the $M(<\theta_E)$ analytical mass estimate for systems with a similar lensing configuration as \clustername\ (azimuthal arc coverage of $\sim40\%$) is biased high by 10\% and has 9\% uncertainty after correcting this bias. This bias is primarily a result of mass distribution ellipticity deviating from the assumed spherical symmetry.  Applying this correction to the analytical estimate results in $M_{corr}(<\theta_E)= 1.07 \pm 0.11 \times 10^{13} M_{\odot}$. We report that our lens model-based mass measurement and the analytical estimate for $M(<\theta_E)$ are consistent within uncertainties. 

\section{Discussion} \label{sec:Discussion}

The strong gravitational lensing model for \clustername\ facilitates the ongoing and future investigation of the lensed $z =\zarc$ background galaxy and its intrinsic properties, such as star formation rate, stellar mass, morphology, and the sizes of star-forming clumps.

\subsection{Comparison to the Previously Published Lensing Analysis of \clustername} \label{sec:Comparison-model}
One of the primary results of this paper is the significant difference between our lens model and the \citetalias{Khullar_2021}\ model. Due to the ground-based seeing limitations, the data used by \citetalias{Khullar_2021}\ did not have the resolution needed to properly resolve the clumpiness of the arc. As a result, their ability to correctly interpret the lensing configuration was reduced, and the low number of available constraints limited the number of free parameters and complexity of the lens model. \autoref{sec:Lensing Constraints} details the differences in modeling assumptions and lensing constraints between the two models.
Because the critical curve was constrained to intersect the arc in three places, the new model predicts that a large portion of the arc forms along the critical curve, resembling a segment of an Einstein ring. As a result, our measurement of the average magnification of the arc is significantly higher than the previous model, $<\mu>=$\magnificationarc\ vs $<\mu>=32^{+8}_{-5}$. This difference, of a factor of $2.4^{+1.4}_{-0.7}$ higher, propagates to the measurements of the intrinsic properties of the lensed galaxy as we describe in \autoref{sec:source}. 
This is perhaps an anecdotal indication of a global concern with the inaccuracy of ground-based data for lensing models.
It is also notable that our model results in a higher fractional statistical uncertainty on the magnification, which is counter intuitive to what is expected from models using space-based resolution. The high statistical uncertainty is due to the proximity of a large portion of the arc to the critical curve, where uncertainties are high. A future analysis of JWST NIRCam data \citep[JWST ID 2566,][]{Khullar2021jwstprop} may reveal infrared color variation that can be used to increase the constraining power of the lensing evidence, reduce the magnification uncertainties, and improve the accuracy of the lens model.

\subsection{Comparison to Strong Lensing Clusters at $z\gtrsim1$}\label{sec:Discussion-mass}
Attempting to compare our mass estimate of the foreground cluster that lenses \clustername\ to that of other $z \gtrsim 1 $ clusters is not straightforward. While numerous clusters and protoclusters are now known in this redshift bin, an estimate of their masses usually comes from mass proxies such as X-ray emission, the Sunyaev Zel'dovich (SZ) effect, or from weak lensing. As such, most mass estimates available in the literature are reported for $r_{200c}$ or $r_{500c}$, which represent the radius from the center of the cluster where the density is 200 or 500 times the critical density of the Universe at the cluster redshift (or similarly, 200 times the mean density, for $M_{200,m}$). Given that our mass measurement relies solely on constraints from the arc and its counter image at the innermost core of the cluster, we cannot reliably extrapolate it beyond a few tens of arcseconds in projection from the BCG. The aforementioned mass proxies lack the spatial resolution to measure the mass at the small cluster-centric radii that strong lensing probes. 
Since a mass comparison to the general population of clusters would be unreliable, we situate the strong lensing inferred mass we measured at the core of the cluster lens \clustername\ in the context of other known strong lensing clusters at $z \gtrsim 1 $.

To date, there are only a handful of strong lensing clusters at $z > 1$ in the literature, with cluster mass estimates using a variety of techniques and at different radii.  
SPT-CL\,J0356$-$5337, a $z = 1.0359$ cluster and a major merger candidate, has a projected mass density of $M(<500\,{\rm kpc}) = 4.0 \pm 0.8 \times 10^{14} M_{\odot}$ \citep{Mahler_2020} inferred from strong lensing analysis, and ${M}_{500c}={3.59}_{-0.66}^{+0.59}\times {10}^{14}$ ${M}_{\odot } $ \citep{Bocquet2019} from SZ effect. 
SPT-CL\,J0546$-$5345 at $z = 1.067$, has a total mass of $M_{200} = 1.0^{+0.6}_{-0.4} \times 10^{15} M_\odot$, measured by \cite{Brodwin_2010} by combining optical, dynamical, X-ray, and SZ mass proxies. \cite{Andersson2011} reported $M_{500} = 5.3 \pm 1.3 \times 10^{14} M_\odot$ for this cluster.
IDCS\,J1426.5$+$3508 is at $z = 1.75$ \citep{Gonzales_2012} with a Chandra X-ray mass measurement of $M_{500,L_x} = 3.3 \pm 0.1 \times 10^{14} M_\odot$ \citep{Stanford_2012} and a SZ mass measurement of $M_{200, m} = 4.3 \pm 1.1 \times 10^{14} M_\odot $ \citep[][]{Brodwin_2012}. 
SPT-CL\,J2011$-$5228 is at $z = 1.064$, with $M_{500} = 2.25\pm0.89 \times 10^{14} M_\odot$ \citep{Reichardt2013} from SZ effect. \cite{Collett_2017} report a mass within an Einstein radius of $\theta_E=14\farcs01$ of $M_E \sim 1.6 \times 10^{14} M_\odot $ from a strong lensing analysis of this cluster. Finally, SPT-CL\,J0205$-$5829 at $z=1.322$ has a mass estimate of $M_{500} = 4.8 \pm 0.8 \times10^{14} M_\odot$ from combined X-ray and SZ analysis \citep{Stalder_2013,Bleem_2015}.

As noted above, absent mass proxies at larger cluster-centric radii, an extrapolation of the strong lensing inferred mass of the cluster lens \clustername\ to $R_{200}$, $R_{500}$, or even to a few hundred kpc, is highly inaccurate. We therefore opted to compare its well-constrained inner core mass to that of the other lensing clusters in its redshift bin, listed above. The projected mass density within the Einstein radius of a strong lensing cluster is well constrained from the strong lensing evidence alone, even without a detailed lens model, as was quantitatively demonstrated by \cite{Remolina2021}. 
We adopt the mass within the Einstein radius derived for SPT-CL\,J2011$-$5228 by \cite{Collett_2017}, $M(<\theta_E)=1.6 \times 10^{14} M_{\odot}$. For SPT-CL\,J0356$-$5337, \cite{Mahler_2020} reports an Einstein radius $\theta_E=14''$ for a source at $z=3$, derived from their best-fit lens model. This translates to $M(<14'')=8.6 \pm 0.9 \times 10^{13} M_{\odot}$, assuming 10\% uncertainty. For IDCS\,J1426.5$+$3508, \cite{Gonzales_2012} report an arc radius of $\theta_E=14\farcs2\pm0\farcs2$ and calculate a lower limit for the mass within the Einstein radius of $M(<\theta_E)=6.9\pm0.3 \times 10^{13} M_{\odot}$, after accounting for the unknown redshift of the arc, and correcting for the overestimate associated with this method, which they estimated to be a factor of $1.6$.  
For the two remaining clusters, SPT-CL\,J0546$-$5345 and SPT-CL\,J0205$-$5829, we measured the arc distance using archival HST data from programs GO-12477 and GO-15294. We reduced the data following the same process as described in \autoref{sec:Data}, and inspected the multi-band imaging (F160W, F814W, F606W) for lensing evidence. We find that the prominent arc in SPT-CL\,J0205$-$5829 is projected $20\pm1 ''$ from the BCG. The HST data of SPT-CL\,J0546$-$5345 reveal several arc candidates, one of which was visible in ground-based data \citep{Staniszewski2009}. The geometry of this arc indicates that it is associated with a massive substructure in the west part of the cluster core, and may not be indicative of the mass of the primary cluster core. We therefore opted to use for our analysis the arc with the largest radius that is not likely to be tracing that structure, at a distance of $23\farcs5\pm1''$ northwest of the BCG. A third prominent arc appears in the near IR (F160W) $14''$ northeast of the BCG. None of the arcs in these two clusters have a known spectroscopic redshift measurement in the literature. 
We calculated the mass within the Einstein radii of these two clusters using the equation in \autoref{sec:results}, for two source redshifts spanning the reasonable redshift range for sources to be lensed and visible in the available data, $z_{arc}=3, 6$. We applied the quadratic empirical correction from \cite{Remolina2020} to obtain the corrected masses and their associated uncertainties here: $M_{corr}(<\theta_E)=(1.6\pm 0.3 - 2.7 \pm 0.5) \times 10^{14} M_{\odot}$ and $M_{corr}(<\theta_E)=(1.4\pm 0.2 - 3.1 \pm 0.5) \times 10^{14} M_{\odot}$ for SPT-CL\,J0546$-$5345 and SPT-CL\,J0205$-$5829, respectively.

To summarize, the core mass of the five strong lensing clusters at $z>1$ ranges between $\sim 0.7-1.6 \times 10^{14} M_{\odot} $. The lensing cluster \clustername\ has a significantly smaller Einstein radius and enclosed mass than other strong lensing clusters at its redshift, \massinRarc. This result is consistent with our expectations from the different discovery methods of these clusters. The SPT clusters were detected due to their deep potential well that is responsible for the SZ signal, while \clustername\ was discovered due to the presence of the giant arc irrespective of its cluster mass. While the core of the cluster is consistent with it being a low-mass galaxy cluster or group, the small Einstein radius and corresponding enclosed mass could alternatively be due to a diffuse mass distribution with low concentration. 

Finally, we note that while not advisable, a measurement of the mass out to large radii of a few hundred kpc is technically possible since the mass reconstruction uses a parameterized lens modeling algorithm that assumes a functional form for the cluster halo. The extrapolation of the projected mass density to 500 kpc from the BCG results in \massinfhkpc. We remind the reader that since the lensing evidence all fall within $\sim 50$ kpc of our BCG, it cannot constrain the outskirts of the mass distribution; in fact it is common practice to fix the truncation parameter ($r_{cut}$) of the cluster halo in the lens modeling process for that reason. 
To gain a better understanding of the statistical uncertainties of the lens model, we re-ran the lens model with $r_{cut}$ free, with a prior of $250 < r_{cut} < 1500$ kpc. The results of the free truncation model are consistent with the fiducial lens model. In particular, the mass within the strong lensing regime, where strong lensing evidence is observed, was unaffected. The statistical spread of the projected mass density increased by  $\sim 50\%$ and the lower limit reduced to $\sim2.5\times10^{14} M_\odot$, but it is likely still an underestimate of the true uncertainty.

\subsection{Revised Source Properties}\label{sec:source}
In this section, we use our improved strong lensing model to update the source galaxy measurements of \citetalias{Khullar_2021}\ that depend on lensing magnification: luminosity, SFR, and stellar mass ($M_\star$). Other results from \citetalias{Khullar_2021}\ that are magnification-independent remain unchanged, e.g., metallicity, dust extinction, and specific SFR. 
\citetalias{Khullar_2021}\ used \texttt{galfit} modeling to measure the multi-band photometry of the giant arc, and \texttt{PROSPECTOR} SED-fitting analysis to determine the physical properties of the source. They reported observed (i.e., before any magnification correction) star formation rate ${\rm SFR}_{\text{0-50 Myr}}=913^{+380}_{-280} M_{\odot} $yr$^{-1}$ and stellar mass ${\log}(M_\star/M_{\odot})=11.63^{+0.19}_{-0.24}$.
We followed the methodology described in \citetalias{Khullar_2021}\ to correct these measurements by the lensing magnification, as follows. For each measurement, we drew a random value from its posterior distribution function and divided it by a random value drawn from the distribution of average arc magnifications. We repeated this process 10,000 times to obtain a statistical sampling and derived the magnification-corrected source property and its uncertainties from the resulting combined distribution. The values that we report are the medians of these 10,000 samples with errors given by the difference between that and the 84th or 16th percentile value. To verify that our methodology is consistent with \citetalias{Khullar_2021}, we applied their magnification to their SED results and confirmed that we recover their demagnified results. 

We then corrected the observed SFR and $M_\star$ for the magnification, in two ways. First, as was done in \citetalias{Khullar_2021}, we correct the values by the average magnification of the giant arc (see \autoref{sec:results}).
Applying our magnification, $<\mu_{arc}> =$ \magnificationarc\ to the SED results, we obtain magnification-corrected values of $\log(M_\star/M_{\odot}) = $ \LMStarn\ and ${\rm SFR} = $ \SFRn\ $ M_{\odot} $yr$^{-1}$.
Our stellar mass and SFR estimates are about 2.4 times smaller than the ones found in \citetalias{Khullar_2021}\ ($\log(M_\star/M_{\odot}) = 10.11^{+0.21}_{-0.26}$, ${\rm SFR} = 27^{+13}_{-9} M_{\odot} $yr$^{-1}$), using the same approach. Similarly, the luminosity should be updated by the ratio of the magnifications, or a $\sim0.92$ mag change in the absolute magnitude previously reported, to $M_{UV}=-21.28\pm0.2$. The reason for this discrepancy comes from the revised magnification  (\autoref{sec:Comparison-model}), which scales directly with luminosity, stellar mass, and SFR.  

Second, we corrected the observed SFR and $M_\star$ by the flux-weighted magnification of the arc. The proximity of the giant arc to the critical curve complicates this measurement directly from the arc, as areas with high flux occur near the critical curve where the magnification and its uncertainties are high. Instead, we take advantage of the relatively uniform magnification of the counter image, which is not bisected by a critical curve, and thus its flux-weighted magnification is approximately the same as its average magnification. The flux ratio between the giant arc and its counter image is the average flux-weighted magnification ratio. We proceeded as described below to obtain the flux ratio $f$, and then calculated the flux-weighted magnification of the giant arc as  $<\mu_{c.i.}>\times f$. 

The total observed flux of the arc and the counter image were measured from the HST F814W data. Since lensing is wavelength-invariant, the flux ratio should not depend on the filter choice, but in the wavelength of F814W the counter image is less contaminated by the BCG. We make use of detailed \texttt{galfit} modeling of the field that will be presented in a forthcoming paper by Sierra et al. (in prep) and refer the reader to that paper for details of the \texttt{galfit} modeling. In short, light sources in a field of view of $10\farcs6 \times 9\farcs1$ enclosing the arc and the counter image were modeled as two-dimensional Sersic components, adding components iteratively to the model until the residuals between the data and model were consistent with the noise level. The resulting scene decomposition allowed us to disentangle the arc and the counter image from contamination from foreground sources (e.g., the BCG light, and a blue source near constraint X1). The flux ratio between the giant arc and the counter image is $f = 14 \pm 0.9$; the resulting flux-weighted magnification of the giant arc is $<\mu_{arc,w}>=$\magnificationarcw. Correcting the SFR and $M_\star$ by the flux-weighted magnification of the arc, we obtain ${\rm SFR} = $ \SFRfw\ $ M_{\odot} $yr$^{-1}$ and $\log(M_\star/M_{\odot}) = $ \LMStarfw, 2.6 and 2.8 times smaller than previously reported by \citetalias{Khullar_2021}, respectively.

A revised measurement of the arc SED is beyond the scope of this paper. The shallow, 4-band HST imaging provide superior resolution to the ground-based data, and a \texttt{galfit} analysis of these data will be presented elsewhere (Sierra et al., in prep).
In future work, we will combine the multi-band HST imaging described here with forthcoming JWST imaging and spectroscopy to obtain a clump-by-clump analysis of the lensed source. This future analysis will significantly supersede an HST+ground based measurement of the arc as a whole, as it will extend farther into the IR, and with greater depth and resolution.

As observed by \citetalias{Khullar_2021}, the magnification-corrected stellar mass and star formation rate places \clustername\ within the range of other $4 \leq z < 6$ galaxies from the \cite{Santini_2017} analysis of four Hubble Frontier Fields. 
\cite{Santini_2017} measured the slope and intercept between stellar mass and star formation rate in log-log space, in different redshift bins. They used $2\sigma$ clipping to exclude outliers, and Monte Carlo simulation to correct for Eddington bias. Applying Equation~1 from \cite{Santini_2017} to the measured mass of \clustername\ gives a predicted main sequence star formation rate of $\log(\rm{SFR})_{\rm MS}=2.06\pm0.13$. Our measured SFR is 5$\sigma$ below the Eddington bias-corrected main sequence (MS) for $5 \leq z<6$, although it falls within $1-\sigma$ of the MS in the $4 \leq z<5$ bin. Since \clustername\ falls below the MS, it is consistent with a galaxy that has low levels of SF relative to SFMS galaxies at $z=5$, although whether it is consistent with a post-starbust/quiescent galaxy remains to be seen (with JWST/NIRSpec observations). Further studies imply that higher stellar mass galaxies which fall below the main sequence experience sudden quenching by AGN feedback while lower stellar mass galaxies experience a slow decrease in SFR due to gas exhaustion \citep{Mancuso_2016, Man_2016}. Because \clustername\ falls in the typical stellar mass range for galaxies at $5 \leq z <6$, it is hard to predict which of these, if either, occurred. This invites a series of questions: Did \clustername\ ever look like the dust-enshrouded early star-forming galaxies? Is AGN feedback responsible for the lack of dust in the galaxy or is there evidence of gas exhaustion? Is the clumpiness that we observe the result of a merger or another process? Is there other evidence to suggest AGN feedback? While these answers are outside of the scope of this paper, our reported magnification values and deflection maps facilitate further investigation of these questions.
Our analysis of \clustername\ is an important stepping stone for analysis of a galaxy observed close to the edge of the Epoch of Reionization.

\section{Summary and Future Work}
We used shallow, multi-band HST imaging to revise the lens model of \clustername, a system with a $z=$ \zarc\ galaxy lensed by a $z=$ \zcluster\ cluster into a bright giant arc, first discovered by \citetalias{Khullar_2021}. The high spatial resolution data reveal the clumpiness of the arc, enabling a detailed lens modeling of the system for the first time. We confirm the candidate counter image identified by \citetalias{Khullar_2021}, and concur with their interpretation that the brightest clump on the arc is a result of high magnification under the critical curve. Other assumptions on the lensing configuration from the ground-based analysis, namely that the giant arc is formed by a common three-image arc configuration, appear to be inconsistent with the HST data. From the revised lensing analysis, we report the following.
\begin{itemize}
  \item We measure a projected mass density enclosed within the Einstein radius of the lensing cluster ($\theta_{\rm E}=5\farcs77$ for a source at $z=$\zarc) of \massinRarc, consistent with an analytical estimate assuming spherical symmetry (\autoref{sec:results}). 
  \item We compile information on Einstein radii and enclosed mass of the five other known strong lensing clusters at the same redshift range, either from the literature, or estimate them directly from archival HST data. We find that the Einstein radius, as well as the enclosed mass of the \clustername\ cluster lens, are significantly lower than other known lensing clusters at $z\gtrsim1$.
  \item We measure an average magnification of $<\mu_{arc}>=$ \magnificationarc\ for the giant arc and $<\mu_{c.i.}>=$ \magnificationci\ for the counter image, about a factor of $2.4^{+1.4}_{-0.7}$ higher than the previous analysis of \citetalias{Khullar_2021}.
  \item We measure the flux-weighted average magnification of the giant arc from the observed flux ratio between the arc and its counter image, $<\mu_{arc,w}>=$\magnificationarcw.
  \item We revise the magnification-dependent measurements of \citetalias{Khullar_2021}\ to account for the higher magnification, by sampling from their \texttt{PROSPECTOR} SED analysis posterior distribution function and dividing by the flux-weighted lensing magnification. We obtain magnification-corrected values of $\log(M_\star/M_{\odot}) =$ \LMStarfw\ and ${\rm SFR} = $ \SFRfw\ $ M_{\odot} $yr$^{-1}$. The luminosity is similarly reduced by the ratio of magnifications, to $M_{UV}=-21.28\pm0.2$.
  \item The stellar mass and star formation rate of the lensed galaxy are within the range of other $5 \leq z <6$ galaxies observed in the Hubble Frontier Fields \citep{Santini_2017}, falling significantly below the $\log(SFR)-\log(M_\star)$ main sequence (corrected for Eddington Bias, \citealt{Santini_2017}).\\
\end{itemize}

Imaging and spectroscopy of \clustername\ with JWST (Cycle~1 program ID 2566, PI Khullar; \citealt{Khullar2021jwstprop}) will be combined with the shallow multi-band HST data presented in this work (HST Cycle~28 GO-16484, PI Stark; \citealt{stark2020hstprop}) to fully analyze the lensed galaxy, on a clump-by-clump basis. These data will also enable a high fidelity measurement of the giant arc and galaxy as a whole, after folding in our new and improved understanding of the lensing configuration of the giant arc. 
In future work, we will revise and improve the lens model by identifying and refining our lensing constraints using the JWST data. We expect that the wider wavelength range will bring out color information to which the HST filters are not sensitive.

\begin{acknowledgments}
This work is based on observations made with the NASA/ESA Hubble Space Telescope, obtained at the Space Telescope Science  Institute, which is operated by the Association of Universities for Research in Astronomy, Inc., under NASA contract NAS 5-26555. These observations are associated with program GO-16484. 
This work also used archival HST data from programs GO-12477 and GO-15294. The HST data presented in this article were obtained from the Mikulski Archive at the Space Telescope Science Institute (MAST). The specific observations analyzed can be accessed via \dataset[DOI]{http://dx.doi.org/10.17909/vk63-hb31}.
Support for HST Program GO-16484 was provided through a grant from the STScI under NASA contract NAS5-26555.
This work used the MATLAB Astronomy and Astrophysics Toolbox \citep[MAAT;][]{Ofek2014}
\end{acknowledgments}

%

\vspace{5mm}
\facilities{HST(ACS, WFC3)}


\software{Source Extractor \citep{bertin1996},
          \LENSTOOL\ \citep{Jullo_2007}, DrizzlePac\footnote{\url{http://www.stsci.edu/scientific-community/software/drizzlepac.html}}, MAAT \citep{Ofek2014}.}





\bibliography{manuscriptv2}{}
\bibliographystyle{aasjournal}



\end{document}